# ON THE CONFIGURATION-LP FOR SCHEDULING ON UNRELATED MACHINES

JOSÉ VERSCHAE* AND ANDREAS WIESE*

ABSTRACT. One of the most important open problems in machine scheduling is the problem of scheduling a set of jobs on unrelated machines to minimize the makespan. The best known approximation algorithm for this problem guarantees an approximation factor of 2. It is known to be $NP$-hard to approximate with a better ratio than $3/2$. Closing this gap has been open for over 20 years.

The best known approximation factors are achieved by LP-based algorithms. The strongest known linear program formulation for the problem is the *configuration-LP*. We show that the configuration-LP has an integrality gap of 2 even for the special case of *unrelated graph balancing*, where each job can be assigned to at most two machines. In particular, our result implies that a large family of cuts does not help to diminish the integrality gap of the canonical *assignment-LP*. Also, we present cases of the problem which can be approximated with a better factor than 2. They constitute valuable insights for constructing an $NP$-hardness reduction which improves the known lower bound.

Very recently Svensson [22] studied the *restricted assignment* case, where each job can only be assigned to a given set of machines on which it has the same processing time. He shows that in this setting the configuration-LP has an integrality gap of $33/17 \approx 1.94$. Hence, our result imply that the unrelated graph balancing case is significantly more complex than the restricted assignment case.

Then we turn to another objective function: maximizing the minimum machine load. For the case that every job can be assigned to at most two machines we give a purely combinatorial 2-approximation algorithm which is best possible, unless $P = NP$. This improves on the computationally costly LP-based $(2 + \varepsilon)$-approximation algorithm by Chakrabarty et al. [6].

## 1. INTRODUCTION

The problem of minimizing the makespan on unrelated machines, usually denoted $R||C_{\max}$, is one of the most prominent and important problems in the area of machine scheduling. In this setting we are given a set of $n$ jobs and a set of $m$ unrelated machines to process the jobs. Each job $j$ requires $p_{i,j} \in \mathbb{N}^+ \cup \{\infty\}$ time units of processing if it is assigned to machine $i$. The scheduler must find an assignment of all jobs to machines with the objective of minimizing the makespan, i. e., the largest completion time of a job.

In a pioneering work, Lenstra, Shmoys, and Tardos [15] give a 2-approximation algorithm based on a natural LP-relaxation. On the other hand, they show that the problem is $NP$-hard to approximate within a better factor than 3/2, unless $P = NP$. Reducing this gap is considered to be one of the most important open questions in the area of machine scheduling [19] and it has been opened for more than 20 years.

Given the apparent difficulty of this problem, people have turned to consider simpler cases. One special case that has drawn a lot of attention is the *restricted assignment* problem. In this setting each job can only be assigned to a subset of machines, and it has the same processing time on all its available machines. That is, the processing times $p_{i,j}$ of a job $j$ equal either a machine-independent processing time $p_j \in \mathbb{N}^+$ or infinity. Surprisingly, the best approximation guarantee known for this problem continues to be the 2-approximation algorithm by Lenstra et al. [15]. However, very recently Svensson [22] shows that the so called *configuration-LP* has an integrality gap of $33/17 \approx 1.94$, and thus is possible to compute in polynomial time a lower bound that is within a factor $33/17 + \varepsilon \approx 1.94 + \varepsilon$ to the optimum. However, no polynomial time algorithm is known to construct a solution with a performance guarantee which is strictly better than 2.

*Technische Universität Berlin, Germany, {verschae,wiese}@math.tu-berlin.de. This work was partially supported by Berlin Mathematical School (BMS) and by DFG Focus Program 1307 within the project "Algorithm Engineering for Real-time Scheduling and Routing".

Another restricted setting that has considered significantly less attention is the *unrelated graph balancing* problem[1]. In this setting each job has finite (possibly different) processing times on at most two machines. Let us remark that the complexity of this case and the *restricted assignment* has different roots. In the *restricted assignment* case, the difficulty is given by the fact that the set of available machines for each job is arbitrary, but its processing time is identical on all these machines. On the other hand, in the *unrelated graph balancing* problem each job has at most two machines available but the processing times may be different on these two machines. Moreover, for instances which have these two properties together, i.e., each job can be assigned to at most two machines on which it has the same processing time, there exists a 1.75-approximation algorithm [8]. In the first part of this paper we focus on the *unrelated graph balancing* problem and show that it is significantly different than the *restricted assignment* problem. We show that the integrality gap of the configuration-LP is 2 in this setting. In contrast, the integrality gap for the restricted assignment case is at most $33/17 < 2$ [22].

In the second part of this paper we consider another related problem that has been in the eyes of the scheduling community in recent years. In the MaxMin-allocation problem we are also given a set of jobs, a set of unrelated machines and processing times $p_{i,j}$ as before. The load of a machine $i$, denoted by $\ell_i$, is the sum of the processing times assigned to machine $i$. The objective is to maximize the minimum load of the machines, i.e., to maximize $\min_i \ell_i$. The idea behind this objective function is a fairness property: Consider that jobs represent resources that must be assigned to machines. Each machine $i$ has a personal valuation of job (resource) $j$, namely $p_{i,j}$. The objective of maximizing the minimum machine load is equivalent to maximizing the total valuation of the machine that receives the least total valuation.

### 1.1. The Minimum Makespan Problem.

**Unrelated machines.** Besides the paper by Lenstra et al. [15] that we have already mentioned, there has not been much progress on how to diminish the approximation gap for $R||C_{\max}$. Shchepin and Vakhania [20] give a more sophisticated rounding for the LP by Lenstra et al. and improve the approximation guarantee to $2 - 1/m$, which is best possible among all rounding algorithms for this LP. On the other hand, Gairing, Monien, and Woclaw [10] propose a more efficient combinatorial 2-approximation algorithm based on unsplittable flow techniques.

In the preemptive version of this problem we are allowed to stop processing a job at an arbitrary time and resume it later, possibly on a different machine. In contrast to the non-preemptive problem, Lawler and Labetoulle [13] show a polynomial time algorithm to compute an optimal preemptive schedule. Thus, it is possible to design an approximation algorithm for $R||C_{\max}$ by using the value of an optimal preemptive schedule as a lower bound. Shmoys and Tardos (cited as a personal communication in [17]), shows that it is possible to obtain a 4-approximation algorithm using this method. More recently, Correa, Skutella and Verschae [7] show that this is best possible by proving that the *power of preemption*, i.e., the worst case ratio between the makespan of an optimal preemptive and non-preemptive schedule, equals 4.

**Restricted Assignment.** The best approximation algorithm for the restricted assignment problem known so far is the $(2 - 1/m)$-approximation algorithm that follows from the more general problem $R||C_{\max}$. As mentioned above, Svensson [22] recently shows how to estimate the optimal makespan within a factor $33/17 + \varepsilon$ in polynomial time. This is done by showing that in this setting the configuration-LP has an integrality gap of at most $33/17$. However, no polynomial time rounding procedure is known.

Other results are known for several special cases, depending on the structure of the set of machines that the jobs can be assigned to, see [16] for a survey. Also, special cases concerning the processing times have been studied. In particular, Lin and Li [18] prove that if all processing times are equal the restricted assignment problem is solvable in polynomial time.

**Graph balancing.** The graph balancing problem can be interpreted as a problem on an undirected graph. The nodes of the graph correspond to machines and the edges correspond to jobs. The endpoints of an edge associated to job $j$ are the machines on which $j$ has finite processing time $p_j \in \mathbb{N}^+$. The objective is to find an orientation of the edges so as to minimize the maximum load of all nodes, where the load of a node is defined as the sum of processing time of its incoming edges (jobs). Notice that the graph may have loops and in that case the corresponding job must be assigned to one particular machine.

[1]To the best of our knowledge, this setting was first mentioned by Ebenlendr, Krčál, and Sgall [8]. However, in that paper they focus on the even more restrictive setting that each job has even the same processing on each of the two machines.

Ebenlendr et al. [8] give a 1.75-approximation algorithm based on an tighter version of the LP-relaxation by Lenstra et al. [15]. They strengthen this LP by adding inequalities that prohibit two large jobs to be simultaneously assigned to a machine. Additionally to the 1.75-approximation algorithm for graph balancing, Ebenlendr et al. [8] also show that it is $NP$-hard to approximate this problem with a factor better than 3/2. This matches the lower bound for the more general problem $R||C_{\max}$. On the other hand, some special cases are studied. For example, it is known that if the underlying graph is a tree, the problem admits a PTAS. If the processing times are either 1 or 2, there is a 3/2-approximation algorithm, which is best possible, unless $P = NP$. For these and more related results see [14] and the references therein.

On the other hand, there is not much known for the unrelated graph balancing problem, where the processing time of a job can be different on its two available machines. To the best of our knowledge, everything known about this problem follows from the more general case $R||C_{\max}$.

1.2. **The MaxMin-Allocation Problem.**
**Unrelated Machines.** The MaxMin-allocation problem has drawn a lot of attention recently. For the general setting of unrelated machines Bansal and Sviridenko [4] show that the configuration-LP has an integrality gap of $\Omega\left(\sqrt{m}\right)$. On the other hand, Asadpour and Saberi [3] show constructively that this is tight up to logarithmic factors and provide an algorithm with approximation ratio $O\left(\sqrt{m}\log^3 m\right)$. Relaxing the bound on the running time, Chakrabarty, Chuzhoy, and Khanna [6] present a poly-logarithmic approximation algorithm that runs in quasi-polynomial time. The best known $NP$-hardness result shows that it is $NP$-hard to approximate the problem within a factor of $2-\varepsilon$ for any $\varepsilon > 0$ [5, 6]. For the special case that there are only two processing times arising in an instance (apart from zero), Golovin [11] gives an $O(\sqrt{n})$-approximation algorithm. He also provides an algorithm that gives at least a $(1-1/k)$ fraction of the machines a load of at least $OPT/k$.
**Restricted Assignment.** Bansal et al. [4] study the case where every job has the same processing time on every machine that it can be assigned to. They show that the configuration-LP has an integrality gap of $O(\log\log m/\log\log\log m)$ in this setting. Based on this they provide an algorithm with the same approximation ratio. The bound on the integrality gap was improved to $O(1)$ by Feige [9] and to 5 and subsequently to 4 by Asadpour, Feige, and Saberi [2, 1]. The former proof is non-constructive using the Lovász Local Lemma, the latter two are given by an (possibly exponential time) local search algorithm. However, Haeupler et al. [12] make the proof by Feige [9] constructive, yielding a polynomial time constant factor approximation algorithm.
**Unrelated Graph balancing.** For the special case that every job can be assigned to at most two machines (but still with possibly different execution times on them) Bateni et al. [5] give a 4-approximation algorithm. Chakrabarty et al. [6] improve this by showing that the configuration-LP has an integrality gap of 2, yielding a $(2+\varepsilon)$-approximation algorithm. Moreover, it is $NP$-hard to approximate even this special case with a better ratio than 2 [5, 6]. In fact, the proofs use only jobs which have the same processing time on their two respective machines. Interestingly, the case that every job can be assigned to at most three machines is essentially equivalent to the general case [5].

1.3. **Our Contribution.** As mentioned before, our main result for the minimum makespan problem is that the configuration-LP has an integrality gap of 2, even in the case of unrelated graph balancing. This implies that any set of cuts that involves only one machine per inequality cannot help to improve the integrality gap of the LP-relaxation of Lenstra et al. [15]. Recall that for the restricted assignment case the configuration-LP has an integrality gap of $33/17 < 2$ [22]. Hence, our result gives an indication that the core complexity of $R||C_{\max}$ lies in the unrelated graph balancing case rather than in the restricted assignment case. In particular, our instances use processing times from the set $\{\varepsilon, 1, \infty\}$. For this case, Svennson [22] proves even an upper bound of $5/3+\varepsilon$ for the integrality gap of the configuration-LP for the restricted assignment problem.

Additionally, we study special cases for which we obtain better approximation factors than 2. In particular, we obtain a $1+5/6$ approximation guarantee for the special case of $R||C_{\max}$ where the processing times belong to the set $[\gamma, 3\gamma] \cup \{\infty\}$ for some $\gamma > 0$. Note that the strongest known $NP$-hardness reductions create instances with this property. Moreover, we show that there exists a $(2 - g/p_{\max})$-approximation algorithm, where $g$ denotes the greatest common divisor of the processing times, and $p_{\max}$ the largest finite processing time. This result generalizes the result by Lin et al. [18], that says that the case where the processing times are either 1 or infinity is polynomially solvable.

We also give a $5/3$-approximation algorithm for the case that an optimal solution assigns only a constant number of jobs to machines where they need more processing time than a $2/3$ fraction of the makespan. We achieve the same approximation guarantee for the case that for all but $O(\log n)$ machines it is priorly known whether they execute such big jobs. These results yield necessary properties for an $NP$-hardness reduction which shows a non-approximability of 2 for $R||C_{\max}$.

We also consider restricted cases of the MaxMin-allocation problem. Our main result for this problem is in the unrelated graph balancing setting, for which we present a simple purely combinatorial algorithm with quadratic running time which has a performance guarantee of 2. This improves on the LP-based $(2+\varepsilon)$-approximation algorithm by Chakrabarty et al. [6]. Their algorithm resorts to the ellipsoid method to approximately solve a linear program with exponentially many variables where the separation problem of the dual is the KNAPSACK problem and can only be solved approximately. Our algorithm is significantly simpler to implement and moreover best possible, unless $P = NP$. Finally, we study what is achievable by allowing *half-integral* solutions, that is, solutions where we allow each job to be split into two halves. We give a polynomial time algorithm that computes a half integral solution whose objective value is within a factor of 2 of the optimal integral solution. Moreover, by loosing an extra factor of 2 in the cost we can transform this solution to a solution with at most $m/2$ fractional jobs. This result contrasts the integral version of the problem for which only an $O(\sqrt{m}\log^3 m)$-approximation algorithm is known.

## 2. LP-BASED APPROACHES

In this section we revise the classical rounding procedure by Lenstra et al. [15] and elaborate on the implications of our results. In the sequel we denote by $J$ the set of jobs and $M$ the set of machines of a given instance.

**The Natural LP-Relaxation.** The natural IP-formulation used by Lenstra et al. [15] uses assignment variables $x_{i,j} \in \{0,1\}$ that denote whether job $j$ is assigned to machine $i$. This formulation, which we denote by LST-IP, takes a target value for the makespan $T$ (which will be determined later by a binary search) and does not use any objective function.

$$\sum_{i\in M} x_{i,j} = 1 \qquad \text{for all } j \in J, \tag{2.1}$$

$$\sum_{j\in J} p_{i,j}x_{i,j} \le T \qquad \text{for all } i \in M, \tag{2.2}$$

$$x_{i,j} = 0 \qquad \text{for all } i,j : p_{i,j} > T, \tag{2.3}$$

$$x_{i,j} \in \{0,1\} \qquad \text{for all } i \in M, j \in J. \tag{2.4}$$

The corresponding LP-relaxation of this IP, which we denote by LST-LP, can be obtained by replacing the integrality condition by $x_{i,j} \ge 0$. Let $C_{LP}$ be the smallest integer value of $T$ so that LST-LP is feasible, and let $C^*$ be the optimal makespan of our instance (or equivalently, $C^*$ is the smallest target makespan for which LST-IP is feasible). Thus, since the LP is feasible for $T = C^*$ we have that $C_{LP}$ is a lower bound on $C^*$. Moreover, we can easily find $C_{LP}$ in polynomial time with a binary search procedure.

Lenstra et al. [15] give a rounding procedure that takes a feasible solution of LST-LP with target makespan $T$ and returns an integral solution with makespan at most $2T$. By taking $T = C_{LP} \le C^*$ this yields a 2-approximation algorithm. The rounding, which we call *LST-rounding*, consists in interpreting the $x_{i,j}$ variables as a fractional matching in a bipartite graph, and then rounding this fractional matching to find an integral solution. The rounding procedure we refer here is a refinement of the original one, derived by Shmoys and Tardos [21] for the *generalized assignment* problem.

**Theorem 1** ([21]). *Let $(x_{i,j})_{j\in J,i\in M}$ be a feasible solution of LST-LP with target makespan $T$. Then, there exists a polynomial time rounding procedure that computes a binary solution $\{\bar{x}_{i,j}\}_{j\in J,i\in M}$ satisfying Equation* (2.1) *and*

$$\sum_{j\in J} \bar{x}_{i,j} p_{i,j} \le T + \max\{p_{i,j} : j \in J \text{ and } x_{i,j} > 0\} \qquad \text{for all } i \in M.$$

By noting that $\max\{p_{i,j} : j \in J \text{ and } x_{i,j} > 0\} \le T$, the previous theorem yields that the rounding procedure embedded in a binary search framework is a 2-approximation algorithm for the makespan problem on unrelated machines.

**Integrality gaps and the configuration-LP.** Shmoys and Tardos [21] implicitly show that the rounding just given is best possible by means of the *integrality gap* of LST-LP. For an instance $I$ of $R||C_{\max}$, let $C_{LP}(I)$ be the smallest integer value of $T$ so that LST-LP is feasible, and let $C^*(I)$ the minimum makespan of this instance. Then the integrality gap of this LP is defined as $\sup_I C^*(I)/C_{LP}(I)$. It is easy to see that if $C_{LP}$ is used as a lower bound for deriving an approximation algorithm then the integrality gap is the best possible approximation guarantee that we can show. Shmoys and Tardos [21] give an example showing that the the integrality gap of LST-LP is arbitrarily close to 2, and thus the rounding procedure is best possible. This together with Theorem 1 implies that the integrality gap of LST-LP equals 2.

It is natural to ask whether adding a family of cuts can help to obtain a formulation with smaller integrality gap. Indeed, for special cases of our problem it has been shown that adding certain inequalities reduces the integrality gap. In particular, Ebenlendr et al. [8] show that adding the following inequalities to LST-LP yields an integrality gap of at most 1.75 in the graph balancing setting:

$$\sum_{j\in J: p_{i,j} > T/2} x_{i,j} \le 1 \qquad \text{for all } i \in M. \tag{2.5}$$

In this paper we study whether it is possible to add similar cuts to strengthen the LP for the unrelated graph balancing problem or for the general case of $R||C_{\max}$. For this we consider the so called *configuration-LP*, defined as follows. Let $T$ be a target makespan, and define $\mathcal{C}_i(T)$ as the collection of all subsets of jobs with total processing time at most $T$, i.e.,

$$\mathcal{C}_i(T) := \left\{ C \subseteq J : \sum_{j\in C} p_{i,j} \le T \right\}.$$

We introduce a binary variable $y_{i,C}$ for all $i \in M$ and $C \in \mathcal{C}_i(T)$, representing whether the jobs assigned to machine $i$ equal to the jobs in $C$. The configuration-LP is defined as follows:

$$\sum_{C\in\mathcal{C}_i(T)} y_{i,C} = 1 \qquad \text{for all } i \in M,$$

$$\sum_{i\in M} \sum_{C\in\mathcal{C}_i(T): C\ni j} y_{i,C} = 1 \qquad \text{for all } j \in J,$$

$$y_{i,C} \ge 0 \qquad \text{for all } i \in M, C \in \mathcal{C}_i(T).$$

It is not hard to see that an integral version of this LP is a formulation for $R||C_{\max}$. Also notice that the configuration-LP suffers from an exponential number of variables, and thus it is not possible to solve it directly in polynomial time. However, it is easy to show that the separation problem of the dual corresponds to an instance of KNAPSACK and thus we can solve the LP approximately in polynomial time. More precisely, given a target makespan $T$ there is a polynomial time algorithm that either asserts that the configuration-LP is infeasible or computes a solution which uses only configurations whose makespan is at most $(1+\varepsilon)T$, for any constant $\varepsilon > 0$ [22]. The following result, which will be proven in the next section, shows that the integrality gap of this formulation is as large as the integrality gap of LST-LP even for the unrelated graph balancing case.

**Theorem 2.** *The integrality gap of the configuration-LP is 2 for the unrelated graph balancing problem.*

Notice that a solution $(y_{i,C})$ of the configuration-LP yields a feasible solution to LST-LP with the same target makespan by using the following formula

$$x_{i,j} = \sum_{C \in \mathcal{C}_i(T): C \ni j} y_{i,C} \qquad \text{for all } i \in M, j \in J. \tag{2.6}$$

This implies that the integrality gap of the configuration-LP is smaller than the integrality gap of LST-LP, and thus it is at most 2. On the other hand, there are solutions to LST-LP that do not yield feasible solutions to the configuration-LP. For example, consider an instance with three jobs and two machines, where $p_{i,j} = 1$ for all jobs $j$ and machines $i$. If we have a target makespan $T = 3/2$, it is easy to see that LST-LP is feasible, but the solution space of the configuration-LP is empty for any $T < 2$.

In the sequel we elaborate on the relation of the two LPs, by giving a formulation in the space with $x_{i,j}$ variables that is equivalent to the configuration-LP. The proof of the following proposition can be found in Appendix A.1. For any set $S \in \mathbb{R}^n$ we define $\mathrm{conv}\{S\}$ to be its convex closure.

**Proposition 3.** *Let $x^C \in \{0,1\}^J$ be the characteristic vector of a configuration $C \in \mathcal{C}_i(T)$, i.e., $x^C_j$ is one if $j \in C$ and zero otherwise. The feasibility of the configuration-LP is equivalent to the feasibility of the linear program defined by Equations* (2.1) *and*

$$(x_{i,j})_{j \in J} \in \mathrm{conv}\{x^C : C \in \mathcal{C}_i(T)\} \qquad \textit{for all } i \in M. \tag{2.7}$$

The last proposition implies that adding any family of cuts to LST-LP that does not remove any vector of the form $(x^{C_1}, \ldots, x^{C_m}) \in \mathbb{R}^{n \cdot m}$, where $C_i \in \mathcal{C}_i(T)$, cannot help to reduce the integrality gap of the linear relaxation. As an example of the implications of this proposition, we note that that adding the cuts given by Inequality (2.5) does not help diminishing the integrality gap of LST-LP for unrelated graph balancing. The same argument shows that the following generalization of these cuts do not help to diminish the gap either: $\sum_{j: p_{i,j} > T/k} x_{i,j} \leq k$ for each machine $i$ and each $k \in \mathbb{N}$.

As another example, we introduce another set of cuts representing the following fact: Given a subset of jobs $S$ with total processing time larger than $T$ on a given machine $i$, at least one job in $S$ must be processed on another machine. Therefore, it must hold that $\sum_{j \in S} x_{i,j} \leq |S| - 1$. More generally, recall that a job can only be assigned to machine $i$ if $p_{i,j} \leq T$. Then, if the total processing time of $S$ on $i$ is larger than $\alpha T$ for some $\alpha \in \mathbb{N}^+$, then at least $\alpha$ jobs in $S$ must be processed elsewhere, i.e. $\sum_{j \in S} x_{ij} \leq |S| - \alpha$. If we strengthen LST-LP with these cuts, we get a linear program whose set of feasible solutions contains the whole polytope defined by expressions (2.1) and (2.7). Proposition 3 together with Theorem 2 implies that these inequalities does not help improve the integrality gap of LST-LP.

## 3. The Configuration-LP

We have seen in the previous section that the configuration-LP implicitly contains a vast class of linear cuts. Hence, it is at least as strong (in terms of its integrality gap) as any linear program that contains any subset of these cuts. However, in this section we prove that the configuration-LP has an integrality gap of 2, even for the special case of unrelated graph balancing. This implies that even all the cuts that are contained in the configuration-LP are not enough to construct an algorithm with a better approximation factor than 2. This is somewhat surprising: if one additionally requires that each job has the same processing time on its two machines then Sgall et al. [8] implicitly proved that the configuration-LP has an integrality gap between 1.5 and 1.75. This indicates that having jobs with different processing times on different machines makes the problem significantly harder.

### 3.1. Integrality Gap for Unrelated Graph Balancing.

We construct a family of instances $I_k$ such that $p_{i,j} \in \{\frac{1}{k}, 1, \infty\}$ for each machine $i$ and each job $j$ for some integer $k$. We will show that for $I_k$ there is a solution of the configuration-LP which uses only configurations with makespan $1 + \frac{1}{k}$. However, every integral solution for $I_k$ has a makespan of at least $2 - \frac{1}{k}$.

Let $k \in \mathbb{N}$ and let $N$ be the smallest integer satisfying $\left(\frac{k}{k-1}\right)^N \frac{1}{k-1} \geq \frac{1}{2}$. Consider two $k$-ary trees of height $N-1$, i.e., two trees of height $N-1$ in which apart from the leaves every vertex has $k$ children. For every leaf $v$, we introduce another vertex $v'$ and $k$ edges between $v$ and $v'$. (Hence, $v$ is no longer a leaf.) Hence, the resulting "tree" has height $N$.

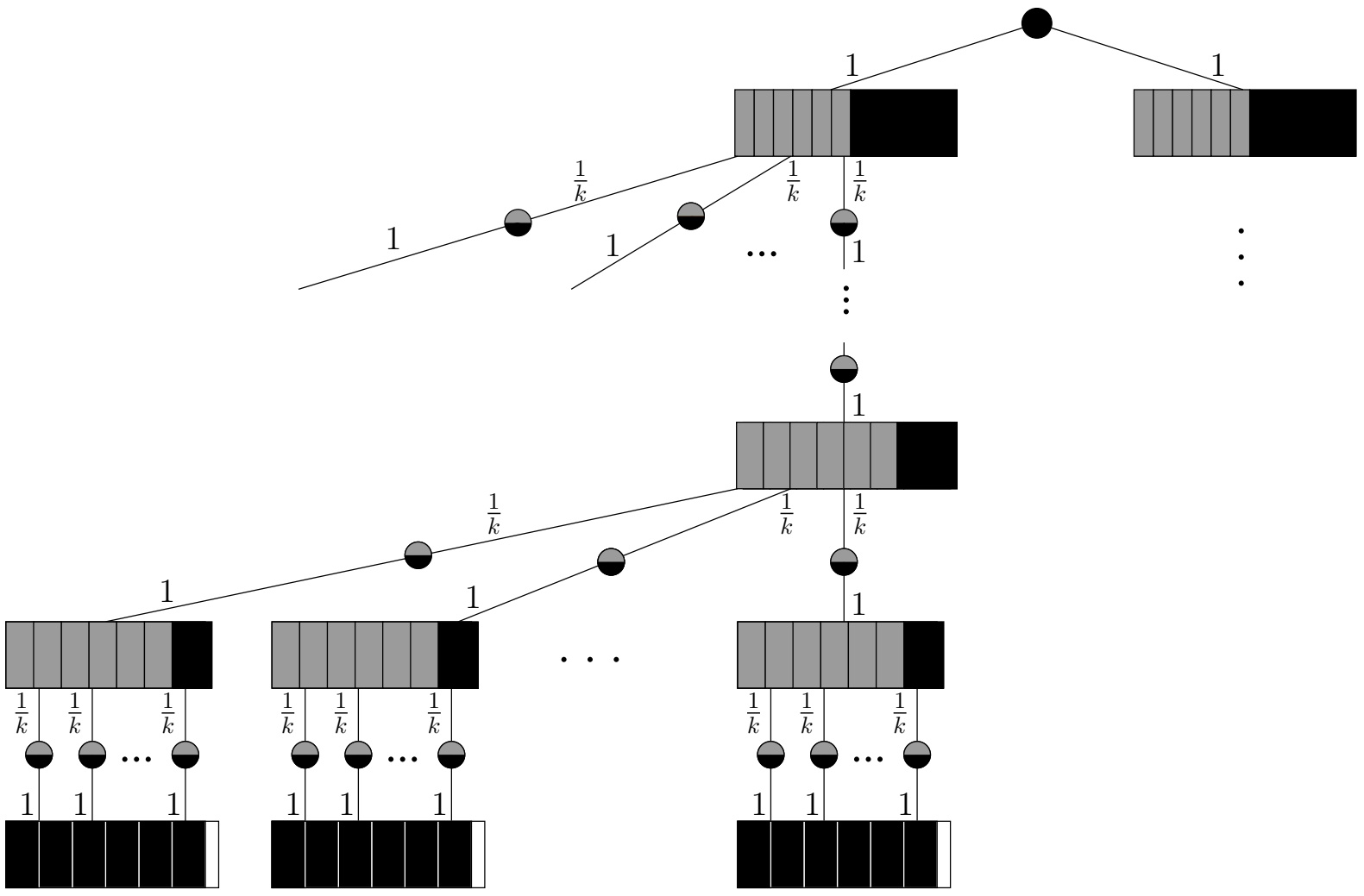

FIGURE 3.1. A sketch of the construction for the instance of unrelated graph balancing with an integrality gap of $2 - O(\frac{1}{k})$. The jobs on the machines correspond to the fractional solution of the configuration-LP for this instances with $T = 1 + \frac{1}{k}$.

Based on this, we describe our instance of unrelated graph balancing. For each vertex $v$ we introduce a machine $m_v$. For each edge $e = \{u, v\}$ we introduce a job $j_e$. Assume that $u$ is closer to the root than $v$. We define that $j_e$ has processing time $\frac{1}{k}$ on machine $m_u$, processing time 1 on machine $m_v$, and infinite processing time on any other machine. Finally, let $m_r^{(1)}$ and $m_r^{(2)}$ denote the two machines corresponding to the two root vertices. We introduce a job $j_{big}$ which has processing time 1 on $m_r^{(1)}$ and $m_r^{(2)}$. Denote by $I_k$ the resulting instance. See Figure 3.1 for a sketch.

As mentioned before, we claim that any integral solution for $I_k$ has a makespan of at least $2 - \frac{1}{k}$. We prove this in the following lemma.

**Lemma 4.** *Any integral solution for $I_k$ has a makespan of at least* $2 - \frac{1}{k}$.

*Proof.* We can assume w.l.o.g. that job $j_{big}$ is assigned to machine $m_r^{(1)}$. If the makespan of the whole schedule is less than 2 then there must be at least one job which has processing time $\frac{1}{k}$ on $m_r^{(1)}$ which is *not* assigned to $m_r^{(1)}$ but to some other machine $m$. We can apply the same argumentation to machine $m$. Iterating the argument shows that there must be a leaf $v$ such that machine $m_v$ has a job with processing time 1 assigned to it. Hence, either $m_v$ has a load of at least $2 - \frac{1}{k}$ or machine $m_{v'}$ has a load of at least 2.

□

Now we want to show that there is a feasible solution of the configuration-LP for $I_k$ which uses only configurations with makespan $1 + \frac{1}{k}$. To this end, we introduce the concept of *$j$-$\alpha$-solutions* for the configuration-LP. A $j$-$\alpha$-solution is a solution for the configuration-LP whose right hand side is modified as follows: job $j$ does not need to be fully assigned but only to an extend of $\alpha \leq 1$.

For any $h \in \mathbb{N}$ denote by $I_k^{(h)}$ a subinstance of $I_k$ defined as follows: Take a vertex $v$ of height $h$ and consider the subtree $T(v)$ rooted at $v$. For the subinstance $I_k^{(h)}$ we take all machines and jobs which correspond to vertices and edges in $T(v)$. (Note that since our construction is symmetric it does not matter which vertex of height $h$ we take.) Additionally, we take the job which has processing time 1 on $m_v$. We denote the latter by $j^{(h)}$.

We prove inductively that there are $j^{(h)}$-$\alpha^{(h)}$-solutions for the subinstances $I_k^{(h)}$ for values $\alpha^{(h)}$ which depend only on $h$. These values $\alpha^{(h)}$ increase for increasing $h$. The important point is that $\alpha^{(N)} \geq \frac{1}{2}$. Hence, there are solutions for the configuration-LP which distribute $j_{big}$ on the two machines $m_r^{(1)}$ and $m_r^{(2)}$ (which correspond to the two root vertices).

The following lemma gives the base case of the induction.

**Lemma 5.** *There is a $j^{(1)}$-$\frac{1}{k-1}$-solution for the configuration-LP for $I_k^{(1)}$ which uses only configurations with makespan at most $1+\frac{1}{k}$.*

*Proof.* Let $m_v$ be the machine in $I_k^{(1)}$ which corresponds to the root of $I_k^{(1)}$. Similarly, let $m_{v'}$ denote the machine which corresponds to the leaf $v'$. For $\ell \in \{1,...,k\}$ let $j_\ell^{(0)}$ be the jobs which have processing time 1 on $m_{v'}$ and processing time $\frac{1}{k}$ on $m_v$.

For $m_{v'}$ the configurations with makespan at most $1+\frac{1}{k}$ are $C_\ell := \left\{j_\ell^{(0)}\right\}$ for each $\ell \in \{1,...,k\}$. We define $y_{m_{v'},C_\ell} := \frac{1}{k}$ for each $\ell$. Hence, for each job $j_\ell^{(0)}$ a fraction of $\frac{k-1}{k}$ remains unassigned. For machine $m_v$ there are the following (maximal) configurations: $C_{small} := \left\{j_1^{(0)},...,j_k^{(0)}\right\}$ and $C_{big}^\ell := \left\{j^{(1)}, j_\ell^{(0)}\right\}$ for each $\ell \in \{1,...,k\}$. We define $y_{m_v,C_{big}^\ell} := \frac{1}{k(k-1)}$ for each $\ell$ and $y_{m_v,C_{small}} := 1-\frac{1}{k-1}$. This assigns each job $j_\ell^{(0)}$ completely and job $j^{(1)}$ to an extend of $k \cdot \frac{1}{k(k-1)} = \frac{1}{k-1}$. □

After having proven the base case, the following lemma yields the inductive step.

**Lemma 6.** *Assume that there is a $j^{(n)}$-$\left(\frac{1}{k-1}\left(\frac{k}{k-1}\right)^n\right)$-solution for the configuration-LP for $I_k^{(n)}$ which uses only configurations with makespan at most $1+\frac{1}{k}$. Then, there is a $j^{(n+1)}$-$\left(\frac{1}{k-1}\left(\frac{k}{k-1}\right)^{n+1}\right)$-solution for the configuration-LP for $I_k^{(n+1)}$ which uses only configurations with makespan at most $1+\frac{1}{k}$.*

*Proof.* Note that $I_k^{(n+1)}$ consists of $k$ copies of $I_k^{(n)}$, one additional machine and one additional job. Denote by $m_v$ the additional machine (which forms the "root" of $I_k^{(n+1)}$). Recall that $j^{(n+1)}$ is the (additional) job that can be assigned to $m_v$ but to no other machine in $I_k^{(n+1)}$. For $\ell \in \{1,...,k\}$ let $j_\ell^{(n)}$ be the jobs which have processing time $\frac{1}{k}$ on $m_v$.

Inside of the copies of $I_k^{(n)}$ we use the solution defined in the induction hypothesis. Hence, each job $j_\ell^{(n)}$ is already assigned to an extend of $\frac{1}{k-1}\left(\frac{k}{k-1}\right)^n$. Like in Lemma 5 the (maximal) configurations for $m_v$ are $C_{small} := \left\{j_1^{(n)},...,j_k^{(n)}\right\}$ and $C_{big}^\ell := \left\{j^{(n+1)}, j_\ell^{(n)}\right\}$ for each $\ell \in \{1,...,k\}$. We define $y_{m_v,C_{big}^\ell} := \frac{1}{k-1}\left(\frac{k}{k-1}\right)^n \frac{1}{k-1}$ for each $\ell$ and $y_{m_v,C_{small}} := 1-\frac{k}{k-1}\left(\frac{k}{k-1}\right)^n \frac{1}{k-1}$. This assigns each job $j_\ell^{(n)}$ completely and the job $j^{(n+1)}$ to an extend of $k \cdot \frac{1}{k-1}\left(\frac{k}{k-1}\right)^n \frac{1}{k-1} = \frac{1}{k-1}\left(\frac{k}{k-1}\right)^{n+1}$. □

Now our main theorem, which we restate here, follows from the previous lemmas.

**Theorem 2.** *The integrality gap of the configuration-LP is 2 for the unrelated graph balancing problem.*

*Proof.* Due to the above reasoning and the choice of $N$ for each of the two subinstances $I_k^{(N)}$ there are $j_{big}$-$\frac{1}{2}$-solutions. Hence, there is a solution for the configuration-LP using only configurations with makespan at most $1+\frac{1}{k}$. With Lemmas 4 this implies that for the instance $I_k$ the integrality gap of the configuration-LP is at least $(2-\frac{1}{k})/(1+\frac{1}{k})$. The claim follows by choosing $k$ arbitrarily large. □

## 4. Cases with Better Approximation Factors than 2

It has been open for a long time whether the approximation factor of 2 [15] for $R||C_{\max}$ can be improved. Our results from Section 3 can be seen as an indicator that this is not possible unless $P = NP$. In this section we identify classes of instances for which a better approximation factor than 2 is possible. This can be understood as a guideline of properties that a $NP$-hardness reduction must fulfill to rule out a better approximation factor than 2.

The inapproximability results for $R||C_{\max}$ given in [8, 15] use only jobs such that $p_{i,j} \in \{1,2,3,\infty\}$. We show now that for classes of instances which use only a finite set of processing times, there exists an approximation algorithm with a performance guarantee which is strictly better than 2. This implies that $NP$-hardness reductions which rule out approximation algorithms with a ratio of $2-\varepsilon$ need an infinite set of processing times for the jobs.

**Theorem 7.** *There exists a $(2-\alpha)$-approximation algorithm for the problem of minimizing makespan on unrelated machines, where $\alpha = \gcd\{p_{i,j}|i \in M, j \in J, p_{i,j} < \infty\}/\max\{p_{i,j}|i \in M, j \in J, p_{i,j} < \infty\}$.*

*Proof.* We follow similar lines as the 2-approximation algorithm by Lenstra et al. [15]. Let $g := \gcd\{p_{i,j}|i \in M, j \in J, p_{i,j} < \infty\}$ and $M := \max\{p_{i,j}|i \in M, j \in J, p_{i,j} < \infty\}$. Note that the optimal makespan of our instance is a multiple of $g$, and therefore we can restrict our target makespan $T$ to be of the form $k \cdot g$ with $k \in \mathbb{N}$. Let $T^*$ be the target makespan defined as the smallest multiple of $g$ that yields a feasible solution to LST-LP. Note that $T^*$ can be found by a binary search procedure. Assume we have computed a fractional solution for LST-LP with target makespan $T^*$. We apply LST-rounding to this fractional solution, obtaining a schedule with load $\ell_i$ on each machine $i$. With basically the same argument as in the proof of Theorem 1, it is easy to see that $\ell_i < T^* + M$. Since $\ell_i$, $M$ and $T^*$ are multiples of $g$, we conclude that $\ell_i \leq T^* + M - g$. A simple calculation then shows the claimed approximation guarantee. □

Now we show that if the execution times of the jobs differ by at most a factor of three then the configuration-LP has an integrality gap of at most $1 + \frac{5}{6} \approx 1.83$. Hence, using reductions of this type one cannot rule out a $2-\varepsilon$ approximation algorithm.

**Theorem 8.** *Consider an instance of $R||C_{\max}$ with a value $\gamma$ such that $p_{i,j} \in [\gamma, 3\gamma] \cup \{\infty\}$ for all machines $i$ and all jobs $j$. Then for this instance the configuration-LP has an integrality gap of at most $1 + \frac{5}{6} \approx 1.83$.*

*Proof.* Assume we are given a value $T$ such that there is a solution for the configuration-LP that uses only configurations with makespan at most $T$. We interpret such a solution as a fractional assignment by using Equation (2.6) and then perform LST-rounding.

Consider a machine $i$. If $T \geq \frac{18}{5}\gamma$ then due to Theorem 1 the makespan of $i$ is bounded by $\frac{18}{5}\gamma + 3\gamma \leq \left(1+\frac{5}{6}\right)T$. So now assume that $T < \frac{18}{5}\gamma$. As mentioned before, the configuration-LP implicitly contains all cuts which are valid for all integral solutions. In particular, it contains the cuts $\sum_{j:p_{i,j} > \frac{T}{2}} x_{i,j} \leq 1$, $\sum_{j:p_{i,j} > \frac{T}{3}} x_{i,j} \leq 2$, and $\sum_j x_{i,j} \leq 3$. From the proof of Theorem 1 it follows that the integral solution obtained after the rounding obeys these cuts as well. This yields a bound $T + \frac{T}{2} + \frac{T}{3} = T(1+\frac{5}{6})$ for the makespan of $i$. □

Note that the above proof implies that adding the three mentioned cuts to the LST-LP yields a $(1+\frac{5}{6})$-approximation algorithm for $R||C_{\max}$.

Finally, we prove that for an improved $NP$-hardness reduction it is crucial that it is not clear what machines execute big jobs. Here, we call a job *big on machine $i$* if $p_{i,j} \geq \frac{2}{3}OPT$. Formally, assume we are given an instance of $R||C_{\max}$ and assume we know exactly what machines execute a big job. For this setting we give a 5/3-approximation algorithm. The key for this algorithm is to incorporate the information about the big jobs into the linear program LST-LP. Then, a similar rounding procedure as LST-rounding yields a bound of $\frac{5}{3}OPT$. In particular, big machines (i.e., machines which run a big job) can have at most one job which is larger than $\frac{1}{3}OPT$. On these machines, the load of the small jobs can increase by at most $\frac{1}{3}OPT$. Also, the load corresponding to a big job can increase by at most $\frac{1}{3}OPT$ (from $\frac{2}{3}OPT$ to $OPT$). Also, on small machines (i.e., where no job is larger than $\frac{2}{3}OPT$) the makespan can increase by at most $\frac{2}{3}OPT$. This yields a bound of $\frac{5}{3}OPT$ for all machines.

If only for $m - O(\log n)$ the additional information is given then we can enumerate the types the unknown machines in polynomial time. See our technical report [23] for details.

**Theorem 9.** *Consider an instace of $R||C_{\max}$ where we know for all but $O(\log n)$ machines whether in an optimal solution they execute a big job. In this case there exists a $5/3$-approximation algorithm.*

With a similar technique we can obtain the following theorem.

**Theorem 10.** *There is a $5/3$-approximation algorithm for instances of $R||C_{\max}$ with at most a constant number of jobs which are big on some machine.*

## 5. MAXMIN ON UNRELATED MACHINES

In this section we study the MaxMin-allocation problem on unrelated machines. First, we investigate the *MaxMin-balancing* problem, where every job can be assigned to at most two machines (with possibly different processing times on each machine). For this case it is known that the configuration-LP has an integrality gap of 2. However, when allowing only polynomial running time it can only be solved approximately which yields a $(2+\varepsilon)$-approximation algorithm for the overall problem. Also, it requires to solve a linear program with a PTAS for KNAPSACK as a separation oracle. In particular, for small $\varepsilon$ this algorithm needs a lot of time and it is highly non-trivial to implement. Instead, we present here a purely combinatorial 2-approximation algorithm with quadratic running time which is quite easy to implement.

After that we present approximation algorithms which compute 2- and 4-approximate half-integral solutions for the general MaxMin-allocation problem. Recall that for this setting the best known approximation algorithm (which computes integral solutions) has a performance guarantee of $O\left(\sqrt{m}\log^3 m\right)$.

5.1. **2-Approximation for MaxMin-Balancing.** We present our purely combinatorial 2-approximation algorithm for MaxMin-balancing. Let $I$ be an instance of the problem and let $T$ be a positive integer. Our algorithm either finds a solution with value $T/2$ or asserts that there is no solution with value $T$ or larger. With an additional binary search this yields a 2-approximation algorithm. For each machine $i$ denote by $J_i = \{j_{i,1}, j_{i,2}, ...\}$ the list of all jobs which can be assigned to $i$. We partition this set into the sets $A_i \dot\cup B_i$ where $A_i = \{a_{i,1}, a_{i,2}, ...\}$ denotes the jobs in $J_i$ which can be assigned to two machines (machine $i$ and some other machine) and $B_i$ denotes the jobs in $J_i$ which can only be assigned to $i$. We define $A'_i$ to be the set $A_i$ without the job with largest processing time (or one of those jobs in case there is a tie). For any set of jobs $J'$ we define $p(J') := \sum_{j \in J'} p_{i,j}$.

Denote by $p_{i,\ell}$ the processing time of job $a_{i,\ell}$ on machine $i$. We assume that the elements of $A_i$ are ordered non-increasingly by processing time, i.e., $p_{i,\ell} \geq p_{i,\ell+1}$ for all respective values of $\ell$. If there is a machine $i$ such that $p(A_i)+p(B_i) < T$ we output that there is no solution with value $T$ or larger. So now assume that $p(A_i) + p(B_i) \geq T$ for all machines $i$. If there is a machine $i$ such that $p(A'_i) + p(B_i) < T$ (but $p(A_i)+p(B_i) \geq T$) then any solution with value at least $T$ has to assign $a_{i,1}$ to $i$. Hence, we assign $a_{i,1}$ to $i$. This can be understood as moving $a_{i,1}$ from $A_i$ to $B_i$. We rename the remaining jobs in $A_i$ accordingly and update the values $p(A_i)$, $p(A'_i)$, and $p(B_i)$. We do this procedure until either

- there is one machine $i$ such that $p(A_i) + p(B_i) < T$, in this case we output that there i
- for all machines $i$ we have that $p(A'_i) + p(B_i) \geq T$.

We call this phase the *preassignment phase*. If during the preassignment phase the algorithm outputs that that no solution with value $T$ or larger exists, then clearly there can be no such solution.

Now we construct a graph $G$ as follows: For each machine $i$ and each job $a_{i,\ell} \in A_i$ we introduce a vertex $\langle a_{i,\ell}\rangle$. We connect two vertices $\langle a_{i,\ell}\rangle, \langle a_{i',\ell'}\rangle$ if $a_{i,\ell}$ and $a_{i',\ell'}$ represent the same job (but on different machines). Also, for each machine $i$ we introduce an edge between the vertices $\langle a_{i,2k+1}\rangle$ and $\langle a_{i,2k+2}\rangle$ for each respective value $k \geq 0$.

**Lemma 11.** *The graph G is bipartite.*

*Proof.* Since every vertex in $G$ has degree two or less the graph splits into cycles and paths. It remains to show that all cycles have even length. There are two types of edges: edges which connect two vertices $\langle a_{i,\ell}\rangle, \langle a_{i',\ell'}\rangle$ such that $i = i'$ and edges connecting two vertices which correspond to the same job on two different machines. On a cycle, the edges of these two types alternate and hence the graph is bipartite. □

Due to Lemma 11 we can color $G$ with two colors, black and white. Let $i$ be a machine. We assign each job $a_{i,\ell}$ to $i$ if and only if $\langle a_{i,\ell}\rangle$ is black. Also, we assign each job in $B_i$ to $i$. In order to turn the above algorithm into an algorithm for the entire problem an additional binary search is necessary to find the correct value of $T$. With appropriate data structures and a careful implementation the whole algorithm has a running time of $O\left(|I|^2\right)$ where $|I|$ denotes the overall input length in binary encoding, see [23] for details.

**Theorem 12.** *There is a 2-approximation algorithm for the MaxMin-balancing problem with running time $O\left(|I|^2\right)$.*

*Proof.* It remains to prove the approximation ratio. Let $i$ be a machine. We show that the total weight of the jobs assigned to $i$ is at least $p(A'_i)/2+p(B_i)$. For each connected pair of vertices $\langle a_{i,2k+1}\rangle$ , $\langle a_{i,2k+2}\rangle$ we have that either $a_{i,2k+1}$ or $a_{i,2k+2}$ is assigned to $i$. We calculate that $\sum_{k\in\mathbb{N}} p_{i,2k+2} \geq p(A'_i)/2$. Since $p_{i,2k+1} \geq p_{i,2k+2}$ (for all respective values $k$) we conclude that the total weight of the jobs assigned to $i$ is at least $p(A'_i)/2 + p(B_i)$. Since $p(A'_i) + p(B_i) \geq T$ the claim follows. □

5.2. **Half-Integral Solutions.** From Theorem 1 we see that during the LST-rounding the load of a machine can increase by at most $\max\{p_{i,j} : j \in J, x_{i,j} > 0\} \leq T$. Assume that one tries to solve the MaxMin-allocation problem with a similar technique. Then the load of a machine might *decrease* by up to (almost) $\max\{p_{i,j} : j \in J, x_{i,j} > 0\}$. Since the latter term might be as large as $T$, during the rounding a machine might lose almost its entire load. Hence, in contrast to $R||C_{\max}$ this method does not yield a constant factor approximation. However, we can adjust the LST-rounding procedure such that we end up with half-integral solutions, i.e., with a solution such that $x_{i,j} \in \{0, \frac{1}{2}, 1\}$ for all machines $i$ and all jobs $j$. Then, the decrease of the load of each machine will be at most $T/2$. Hence, the value of the solution is by at most a factor of 2 away from the optimal integral solution.

If one is interested in half-integral solutions where only few jobs are distributed on two machines, a technique similar to the one used in the 2-approximation for MaxMin-balancing yields a half-integral solution in which at most $m/2$ jobs are half assigned on two machines. This procedure loses only a factor of 2 in the objective function. We summarize these algorithms in the following theorems.

**Theorem 13.** *There is a polynomial time algorithm that computes half-integral solutions $HALF(I)$ for instances $I$ of the MaxMin-allocation problem such that $HALF(I) \geq \frac{1}{2}OPT(I)$.*

**Theorem 14.** *There is a polynomial time algorithm that computes half-integral solutions $HALF^{(2)}(I)$ for instances $I$ of the MaxMin-allocation problem such that $HALF^{(2)}(I) \geq \frac{1}{4}OPT(I)$ and at most $m/2$ jobs are distributed on two machines.*

5.3. **Tractable Cases.** Similarly as for $R||C_{\max}$ if the number of big jobs is bounded by a constant $c$ or if we knew what machines execute a big job in an optimal solution we can guarantee better approximation factors. Here, we define a job $j$ to be *big on machine $i$* if $p_{i,j} \geq \frac{1}{2}OPT$. With similar techniques as in Section 4 we can prove the following theorems.

**Theorem 15.** *Consider an instance of the MaxMin-allocation where we know for all but $O(\log n)$ machines whether in an optimal solution they have a big job. In this case there exists a 2-approximation algorithm.*

**Theorem 16.** *There is a 2-approximation algorithm for instances of the MaxMin-allocation problem with at most a constant number of jobs which are big on some machine.*

## APPENDIX A. OMITTED PROOF

A.1. **Proof of Proposition 3.** Let $(x_{i,j})_{i\in M,j\in J}$ be a solution satisfying (2.1) and (2.7) for a given $T$. We show that the configuration-LP is feasible for the same value of $T$. Indeed, $(x_{i,j})_{j\in J}$ is a convex combination of vectors in $\{x^C : C \in \mathcal{C}_i(T)\}$, and thus

$$(x_{i,j})_{j\in J} = \sum_{C\in\mathcal{C}_i(T)} y_{i,C} \cdot x^C,$$

for some values $y_{i,C} \geq 0$ such that $\sum_{C\in\mathcal{C}_i(T)} y_{i,C} = 1$. Moreover, for each $j \in J$,

$$1 = \sum_{i\in M} x_{i,j} = \sum_{i\in M} \sum_{C\in\mathcal{C}_i(T)} y_{i,C} \cdot x_j^C = \sum_{i\in M} \sum_{C\in\mathcal{C}_i(T):C\ni j} y_{i,C}.$$

This shows that $(y_{i,C})$ is a solution to the configuration-LP. The converse implication follows from reversing the argument just given.